\documentclass[published]{JHEP}
\JHEP{03(1999)017}
\usepackage{epsfig}

\def\lsim{\mathrel{\mathpalette\fun <}}
\def\gsim{\mathrel{\mathpalette\fun >}}
\def\fun#1#2{\lower3.6pt\vbox{\baselineskip0pt\lineskip.9pt
  \ialign{$\mathsurround=0pt#1\hfil##\hfil$\crcr#2\crcr\sim\crcr}}}
\newcommand{\beq}{\begin{equation}}
\newcommand{\eeq}{\end{equation}}

\title{On the disintegration of cosmic ray nuclei by solar photons}

\author{Luis N. Epele, Silvia Mollerach and Esteban Roulet\\
	Depto. de F\'\i sica, Universidad Nacional de La Plata\\ 
	CC67,  1900, La Plata, Argentina\\ 
	Email: \email{epele@venus.fisica.unlp.edu.ar},
	\email{mollerach@venus.fisica.unlp.edu.ar},
	\email{roulet@venus.fisica.unlp.edu.ar}}

\abstract{We discuss in detail the possibility of observing pairs of
simultaneous parallel air showers produced by the fragments of cosmic
ray nuclei which disintegrated in collisions with solar photons. We
consider scenarios with different cosmic ray compositions, exploring
the predicted rates for existing and planned detectors and looking for
methods to extract information on the initial composition
from the characteristics of the signal. In particular, we find that
fluorescence detectors, such as HiRes or the Telescope Array, due to
their low threshold ($\sim 10^{17}$~eV) and large area ($\sim
10^4$~km$^2$) may observe several events per year if cosmic rays at
those energies are indeed heavy nuclei. The possibility of exploiting
the angular orientation  of the plane containing the two showers to
further constrain the cosmic ray composition is also discussed.}

\received{December 8, 1998}
\accepted{March 23, 1999}
\keywords{High Energy Cosmic Rays, Electromagnetic Processes and Properties}

\begin{document}

\section{Introduction}

A long time ago, Zatsepin and Gerasimova~\cite{za51,ge60} suggested
that the photo-di\-sin\-te\-gra\-tion of cosmic ray (CR) nuclei with solar
photons could allow to study the CR composition at very high energies,
$E\gsim 0.1$~EeV (1~EeV$=10^{18}$~eV). 
Indeed, if a nucleon is stripped from a heavy
nucleus on a photo-disintegration process, the comparison of the
energies of the surviving nucleus with that of the emitted nucleon
would directly provide the mass of the primary nucleus.

In order to be able to observe this kind of events, one needs that the
probability for the disintegration process to take place
 as the nucleus traverses the solar system be non-negligible. It is
also necessary that
the separation among the fragments be sufficiently large so as to allow
their individual identification but however not larger than the
spatial coverage of the CR detectors.

To maximise the photo-disintegration probability, previous works have
focused on CR energies for which the interaction with solar photons
takes place at the giant resonance for photo-disintegration, i.e.\ such
that the typical photon energy in the CR rest frame is 10--30~MeV
(e.g.\ $E\simeq 1$~EeV for iron nuclei).

Regarding the separation among the fragments, 
Gerasimova and Zatsepin (GZ)  originally estimated that it 
was of the order of 1~km~\cite{ge60}, 
based on the splitting resulting
from the transverse momenta acquired by the fragments 
in the disintegration process.
However, as Zatsepin later realised, the dominant contribution to the
splitting actually results from the deflection of the fragments in the
solar system magnetic field (which was initially largely
underestimated). This deflection depends on the charge to mass ratio
of the fragments, and is hence different for nuclei, protons or neutrons. 

\pagebreak[3]
In a recent reanalysis of this mechanism, Medina-Tanco and 
Watson~\cite{me98} have performed detailed orbit integrations for incident Fe
nuclei, using a realistic model for the solar system magnetic
field. With these improved computations, they 
showed that the resulting separations are typically much
larger than 10~km, and hence exceed the typical sizes of existing
detectors, what strongly limits their observability. Going to higher
energies in order to reduce the average shower separation down to an
observable range has however the problem of a much reduced rate, since
not only the cross section decreases outside the giant resonance, but
more importantly the CR flux falls abruptly.  Hence, the conclusion of
that work regarding the observability of the effect was not
encouraging. 

In this work we want to extend previous analyses in several
ways. Since the ultra-high energy CR are most probably not
dominantly Fe nuclei, we study the GZ effect for different initial
compositions. In order to
cover all the range of possible CR nuclei masses we take
as prototypical examples of heavy, intermediate
and light compositions the nuclei 
$^{56}$Fe, $^{16}$O and $^4$He. We find that
although the cross sections are reduced going to lighter nuclei, for a
given initial energy the separation among the fragments are smaller
(approximately proportional to the mass number $A$), and hence this
helps to make this kind of events observable. 
Also, since the energy of the emitted nucleon, which is
smaller than the parent cosmic ray energy by a factor $A$, becomes
higher the lighter is the initial mass composition, this allows for an
easier detection of the nucleonic fragments.\footnote{For a Fe nucleus
with $E\simeq 7\times 10^{17}$~eV, as considered in~\cite{me98}, the
emitted nucleon has $E\simeq 10^{16}$~eV, which is below the threshold
of the large extended air shower arrays.} We have also extended the
cross sections beyond the pion-production threshold in this study.

The determination of the parent CR mass
through the comparison of the energies of the two showers may be
problematic in some cases. This occurs for instance
if the determination of the nucleonic fragment energy is
poor, as would happen if the energy is small so that only one detector
in a ground array is hit by the shower (e.g.\ for a fragment energy 
$\simeq 10^{17}$~eV for  
Auger), or if more than one nucleon is emitted in the
photo-disintegration. One may then attempt to use the
size of the separation among the fragments
to infer the initial composition. However, for a given initial energy
the separation among the fragments also depends on the distance from
the Earth at which the photo-disintegration took place. Hence, 
nuclei with different masses but interacting at different distances
may give rise to equally separated showers.
We have found that this degeneracy can be lifted in many cases by
looking at the inclination of the plane containing the two fragments, 
since this angle depends on the distance
from the Earth at which the photo-disintegration took place. 
Looking then at the whole available information (total energy,
separation and inclination of the fragments) it will be possible to
get further insights into\linebreak the initial CR composition.

\section{The photo-disintegration process}

As a CR traverses the solar system, it encounters a flux of energetic
($\epsilon_\gamma\simeq 1$~eV) solar radiation. 
 The lifetime for photo-disintegration off this radiation, computed
in the rest frame of the CR nucleus (primed quantities), is just
\beq
{1\over \tau'}=c\int_0^\infty {\rm d}\epsilon' {{\rm d}n'\over {\rm
d}\epsilon'} \sigma(\epsilon') \,,
\eeq
with $\sigma(\epsilon')$ the photo-disintegration cross section.
Transforming back to the lab (Earth) frame, using that
$\epsilon'=\epsilon \gamma(1+\beta \cos\alpha)$, where
$\beta=v/c\simeq 1$ in terms of the CR velocity $v$, $\gamma$ is the
usual relativistic factor and $\alpha$ is
the angle between the directions of propagation of the CR and of the
photon, one finds for the CR mean free path
\beq
{1\over \lambda(\xi)}=\int_0^\infty {\rm d}\epsilon {{\rm d} n\over
{\rm d} \epsilon}(\xi)\sigma(\epsilon')[1+\beta\cos\alpha]\,.
\eeq
Here $\xi$ is a coordinate measuring the distance
 from the Earth along the arrival
direction of the CR, so that if $\hat\xi$ is the unit vector in that
direction, the CR velocity is ${\bf v}\simeq
-c\hat{\bf \xi}$ and $\cos\alpha=\hat{\bf \xi}\cdot\hat{\bf r}$,
with $r$ the spherical radial coordinate centered in the~sun.

The probability that an incoming CR suffers fragmentation along its
path towards the Earth is then
\beq
\eta_{GZ}=1-{\rm exp}\left[-\int_0^\infty {\rm d}\xi{1\over
\lambda(\xi)} \right] .
\eeq
This probability turns out to be small ($\ll 10^{-3}$), since
$\lambda(\xi)$ is typically much larger than the characteristic solar
system dimensions. For practical purposes, we will integrate up to
$\xi_{max}=5$~AU, since the contribution from larger distances turns
out to be negligible due to the decreasing photon flux.

The solar photon flux can be obtained as a blackbody spectrum with
the temperature of the solar surface, $T_s=5770^\circ$K ($k_BT_s\simeq
0.5$~eV), normalised so as to reproduce the solar luminosity,
$L_\odot=4\pi r^2c\int{\rm d}\epsilon\ \epsilon
{\rm d}n/{\rm d}\epsilon$. One
then finds 
\beq
{{\rm d}n\over {\rm d}\epsilon}=7.2\times 10^7{\epsilon^2\over {\rm
exp}(\epsilon/k_BT_s)-1} \left({1\ {\rm AU}\over r}\right)^2 [{\rm eV\
cm}]^{-3}\,.
\eeq

Regarding the photo-disintegration cross section $\sigma$, one can
distinguish different regimes according to the photon energy in the
nucleus rest frame. There is first the domain of the giant resonance,
from the disintegration threshold ($\epsilon'\simeq 2$~MeV) up to
$\sim 30$~MeV, in which a collective nuclear mode is excited with the
subsequent emission of one (or possibly two) nucleons. Beyond the
giant resonance and up to the pion production threshold
($\epsilon'\simeq 150$~MeV) the excited nucleus decays dominantly by
two nucleon (quasi-deuteron effect) and multinucleon
emission. Detailed fits to the cross sections and branching fractions
in these first two regimes were performed by Puget, Stecker and
Bredekamp~\cite{pu76} in their study of 
the CR photo-disintegrations. For our
purposes, it will be enough to adopt the simpler expressions for the
total cross section given in~\cite{ka93}, which are
\beq
\sigma(\epsilon')=\sigma_{GR}(\epsilon')\equiv 
1.45 A{(\epsilon' T)^2\over (\epsilon'^2-\epsilon_0^2)^2+
(\epsilon'T)^2} {\rm mb \qquad for\ \epsilon'<30\ MeV},
\eeq
where $T=8$~MeV and $\epsilon_0=42.65 A^{-0.21}$~MeV for $A>4$ and 
$\epsilon_0=0.925 A^{2.433}$~MeV for $A\leq 4$. For higher energies we
take
\beq
\sigma(\epsilon')={\rm max}\left\{ \sigma_{GR}(\epsilon'),A/8\ {\rm
mb}\right\} \qquad {\rm for\ 30\  MeV<\epsilon'<150\ MeV}\,.
\eeq
We do not cutoff $\sigma_{GR}$ above 30~MeV since for light nuclei the
peak of the giant resonance is already close to 30~MeV and the
resonance is wide.

When considering light nuclei and/or very energetic CR
($E>1$~EeV), it is necessary to consider in some
detail the cross section above the pion production threshold. From this
threshold up to $\epsilon'\simeq 600$~MeV, the disintegration is
dominated by the delta production $(\gamma N\to \Delta\to N\pi$, where
$N$ is a nucleon).\footnote{We note that using Rudstam type fits to
the proton-nucleus cross section~\cite{si73,el95} 
to estimate the photo-nuclear rates,
or also photo-disintegration data from bremsstrahlung
photons~\cite{jo73},  one 
misses the strength of the $\Delta$ resonance peak.}
In this regime, the cross section per nucleon is found to be almost
independent of the
target nucleus~\cite{ar81,ah85}. Due to nuclear Fermi motion and
interaction effects, the peak in this cross section is reduced and
the resonance becomes somewhat wider than in the free
proton case. (Higher resonances beyond $\Delta(1232)$ should also
contribute for $\epsilon'\sim 1$~GeV.) 
 
We have parametrised the total photo-absorption cross section
in the region above the pion production threshold as
\beq
\sigma(\epsilon')=A\left[{1\over 8}+S\ \tilde\epsilon\ {\rm exp}
\left({1-\tilde\epsilon^\nu\over\nu}\right)\right] {\rm mb}\,,\qquad {\rm
for}\  \epsilon'>150\ {\rm MeV}\,,
\eeq
where $\tilde\epsilon\equiv(\epsilon'-150\ {\rm MeV})/\epsilon_1$. 
A good fit to the experimental data~\cite{ar81} 
is obtained with $\epsilon_1=
180\ {\rm MeV}$, $S=0.3$ and $\nu=1.8$, so that the maximum in the cross
section ($\sigma_{max}/A=(1/8+S)$~mb$\simeq 0.4$~mb) occurs for an
energy $\epsilon'=\epsilon_1+150$~MeV$\simeq 330$~MeV. The value of  
$\nu$ gives the right size of the resonance width.

Near the resonance peak, the one nucleon emission by direct knock-out
is relevant, but the nucleon multiplicity increases at higher
energies. The pions emitted will decay, for the typical $\gamma$
factors considered ($\gamma\sim 10^7$--$10^9$), in a distance much
shorter than 1~AU, and only their decay products can reach the
atmosphere. However, these decay products will have an
energy much below the initial CR energy and hence will pass largely
unnoticed. Furthermore, the  muons from
$\pi^\pm$ decays will be significantly deflected by the solar system
magnetic fields and hence arrive far from the nucleonic fragments.

\section{The deflection of the fragments}

After the parent CR photo-disintegrates, the fragments travel towards
the Earth  and are deflected
by the solar system magnetic field with respect to
 their initial trajectories. The resulting deflection can be
computed by integrating the transverse displacement due to the Lorentz
force from the fragmentation point, at a distance $\xi$, up to the
Earth.\footnote{In the following we neglect the deflection due to the
transverse momenta acquired by the nucleonic fragments in the
photo-disintegration, since it leads to separations $\lsim
0.2$~km $\sqrt{A/(E/\ {\rm EeV})}(\xi/1$~AU). The corresponding
separation for an emitted alpha particle would be four times
smaller.\label{ft3}}   In
this way one finds
\beq
{\bf d}_f(\xi)={Z_f e\over   A_f m \gamma c}\int_\xi^0{\rm d}\xi'
\int_{\xi'}^0{\rm d}\xi'' {\bf B}(\xi'')\times \hat\xi\,,
\label{eqdf}
\eeq
where $e$ and $m$ are the proton charge and mass. The deflection $d_f$
of each fragment is then proportional to its charge to mass ratio,
$Z_f/A_f$. For the separation among the fragments what matters is the
difference $|Z_1/A_1-Z_2/A_2|$. If the emitted nucleon is a neutron,
this is just $Z/(A-1)$, with $Z$ and $A$ associated to the charge and
mass of the parent nucleus.
If a proton is emitted, the difference is given by
$(A-Z)/(A-1)$. Thus, the separation among the two fragments is
approximately given by eq.~(\ref{eqdf}) but replacing $Z_f/A_f$ by 
1/2 for Fe and O, and by 2/3 for He. In the case of multiple nucleon
emission, all protons will be deflected by the same amount by the
magnetic field (and all neutrons will be undeflected), so that they
will produce essentially one shower (the typical separation of the
strongly overlapped nucleonic showers, see footnote~\ref{ft3}, is smaller than the
separation of the detectors in the largest arrays, which is 
e.g.\ 1.5~km for Auger).
 The interesting possibility also 
appears of having three simultaneous showers caused by the main
fragment, the emitted proton(s) and neutron(s), all contained in the
same plane and equally separated. 

As in~\cite{me98}, we consider for the solar system magnetic field the
realistic model of Akasofu, Gray and Lee~\cite{ak79}, which describes
the field within the heliosphere, taken as a sphere of radius
$r_2=20$~AU.  It consists of four components (written here for 
an odd solar cycle, i.e.\ with the north polar region having
the S magnetic pole):
\begin{itemize}
\item[$a)$] The solar dipole contribution can be written, in
cylindrical coordinates around the sun, as
\begin{eqnarray}
B_z^{dip}&=&{B_sr_1^3\over
2}(\rho^2-2z^2)(z^2+\rho^2)^{-5/2}\,, \\
B_\rho^{dip}&=&-{3B_sr_1^3\over 2}\rho
z(z^2+\rho^2)^{-5/2}\,, \label{bdip}\\
B_\phi^{dip}&=&0\,,
\end{eqnarray}
where $B_sr_1^3/2$ is the magnetic 
dipole moment of the sun. Taking  $r_1$ to
be the solar radius, $r_1=R_\odot$,  one has $B_s=2$~Gauss.

\item[$b)$] The dynamo component arises from the rotation of the
sun in the dipole field. It is generated by a sheet current
distribution that flows outwards along the polar axis, 
 then flows along the
heliosphere towards the ecliptic plane and then radially 
inwards back to the sun.
The resulting field is in the $\phi$ direction,
\beq
B_\phi^{dyn}= {\rm sign}(z) B_{\phi_0} \frac{\rho_0}{\rho}\,,
\qquad r_1 < \sqrt{z^2 + \rho^2} < r_2\,,
\eeq
where $B_{\phi_0}=3.5\times 10^{-5}$~Gauss and $\rho_0=1$~AU.

\item[$c)$] The ring current component arises from a sheet
equatorial current extending up to $r_2$, with current density 
$\propto \rho^{-2}$. The resulting magnetic field is given by
\begin{eqnarray}
B_z^{ring}&=& B_{\rho_0}{\rho_0}^2 \int_0^\infty
{\rm d}k\ k G(k) J_0(k\rho) \exp(-k|z|)\,,\label{bzr}
\\
B_\rho^{ring}&=&{\rm sign}(z) B_{\rho_0}{\rho_0}^2 \int_0^\infty
{\rm d}k\ k G(k) J_1(k\rho) \exp(-k|z|)\,,\label{brr}
\\
B_\phi^{ring}&=&0\,,
\end{eqnarray}
where
$$G(k)=\frac{1}{k} \left[\sqrt{k^2+\frac{1}{r_2^2}}-
\sqrt{k^2+\frac{1}{r_1^2}}
+\frac{1}{r_1}-\frac{1}{r_2}\right],$$ 
 $J_{0,1}$ are Bessel functions and $B_{\rho_0}= -3.5\times 10^{-5}$
Gauss.

For the region of the solar system interesting for the present work
($r_1 \ll r < 5$~AU) the integrals in eqs.~(\ref{bzr}) and (\ref{brr})
can be approximated by taking the limit $r_1 \rightarrow 0$ and 
$r_2 \rightarrow \infty$ in $G(k)$. This gives
\begin{eqnarray}
B_z^{ring}&\simeq& B_{\rho_0}{\rho_0}^2 |z| (z^2+\rho^2)^{-3/2}\,,\label{bzra}
\\
B_\rho^{ring}&\simeq&{\rm sign}(z)
 B_{\rho_0}{\rho_0}^2 \rho (z^2+\rho^2)^{-3/2}\,
.\label{brra}
\end{eqnarray}

\item[$d)$] The sunspot component is approximated by an ensemble of
180 dipoles of radius 0.1~$R_{\odot}$ located in the equatorial plane,
just below the solar surface at a radius of 0.8~$R_{\odot}$
(their expressions are similar to eq.~(\ref{bdip}) with $B_d=1000$
Gauss). They provide a strong magnetic field near the sun surface and
allow the ring current field lines to connect to the solar surface.
\end{itemize}

All the components change sign with the eleven years solar cycle. We plot
in fig.~\ref{f1} the three components of the magnetic field $B_\phi, B_\rho$
and $B_z$ as a function of $\rho$ for three different values of
$z$.
At large distances the dominant contribution is given by $B_\phi$ and
$B_\rho$ (due essentially to the dynamo and ring current components
respectively).\footnote{The $B_z$ component is at variance with the one
plotted
in ref.~\cite{me98}, and is proportional to $r^{-3}$ for large radius. 
However, being this one
the smallest componet, it has little impact on the results.}

We notice that the field model adopted gives a
consistent description  of Parker's spiral magnetic field
configuration, and allows us to discuss the general features of the GZ
events. Some departures from this simple model occur for instance due
to the variation of the overall field normalization during the solar
cycle or from a possible tilt of the magnetic equator with respect to
the ecliptic, and these can be taken into account in the event of an
actual detection. Sporadic magnetic perturbations such as shock waves
produced by solar flares could also add some noise.

\EPSFIGURE[t]{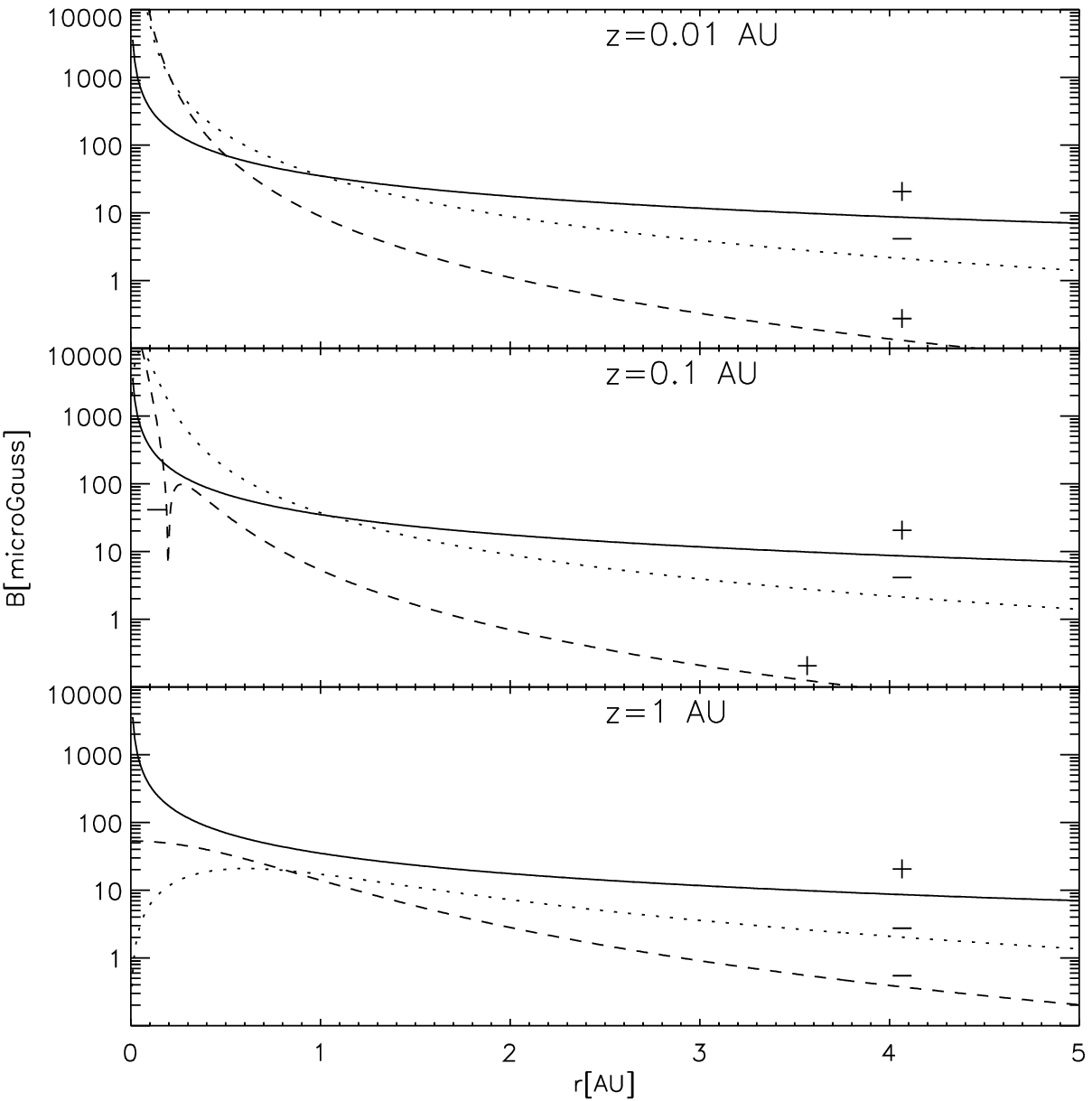}{The three components of the solar system 
magnetic field, $B_\phi$
(solid line), $B_\rho$ (dotted line) and $B_z$ (dashed line), as a
function of the radial cylindrical coordinate $\rho$ for values of the $z$
coordinate above the ecliptic $z = 0.01$~AU, 0.1~AU and 1 AU. We have
plotted the modulus of each component and the sign is indicated in each
curve.\label{f1}}

We describe the CR position by the distance $\xi$ to the Earth and the
latitude and longitude 
angular coordinates $(b, \ell)$, defined analogously to the Galactic
ones, with $b = 0$ in the ecliptic plane and the sun at the origin
of $\ell$, with increasing $\ell$ values to the left (with the N
up). The 
separation among the fragment positions can be decomposed along the 
$b$ and $\ell$ directions, ${\bf d}= d_\ell \hat\ell +d_b \hat b$. 
We will use the
modulus $d \!=\! \sqrt{d_\ell^2 + d_b^2}$ and the angle with respect to the
$\hat\ell$ direction, atan($d_b/d_\ell$), to describe~it. 

We can use the symmetry properties of the magnetic field about the
ecliptic plane to infer the relationship among the separations between
the fragments arriving from directions with the same longitude $\ell$,
but opposite latitude $b$, fragmenting at the same distance from the
Earth (so that the sign of $\xi_z$ changes, but the component of 
${\bf \xi}$ in the ecliptic plane is the same). Under the reflection
$z \rightarrow -z$, the radial and azimuthal components of {\bf B}
change sign, while $B_z$ is invariant. Thus, the $z$ component of the
product ${\bf B}\times \hat\xi$ will change sign, while the
orthogonal component remains unchanged. This means that
$d_b(-b,\ell)=-d_b(b,\ell)$ and $d_\ell(-b,\ell)=d_\ell(b,\ell)$. Or,
in other words, rays arriving from the same $\ell$ but opposite $b$
and produced at the same distance from the Earth have separations
between fragments equal in modulus but with opposite angles with
respect to $\hat\ell$. Hence, when integrating the fragment orbits,
only half of the celestial sphere needs to be evaluated.\footnote{This
symmetry is not apparent in ref.~[3], probably due to insufficient
numerical accuracy.} We also note
that in an even solar cycle all the field components
 are reversed, but this leaves
unchanged both the separations and the angles.

\section{Results}

Typical separations for different parent nuclei are illustrated in
fig.~\ref{f2} for two  different energies.
The mean separation for each arrival
direction is plotted (dashed lines), where the mean 
is weighted with the
fragmentation probability for  the photo-disintegration processes
to occur  at different distances from the Earth, i.e.\
\beq
\langle  d(\ell,b)\rangle = \frac{1}{\eta_{GZ}}
\int^{\xi_{max}}_0 {\rm d}\xi  d(\xi,\ell,b)
\frac{{\rm d}\eta_{GZ}}{{\rm d}\xi}\,,
\eeq
where d$\eta_{GZ}/{\rm d}\xi\simeq 1/\lambda(\xi)$.

For Fe nuclei with $E=1$~EeV most of the separations are larger than
100~km and are hence out of the reach 
of existing detectors, as noticed in ref.~\cite{me98}. 
For lower energies this problem worsens, and
 for instance for 
$E= 0.1$~EeV most of the separations are larger than 
500~km for Fe.  The
situation improves for lighter nuclei, due to the factor $\gamma$ in
eq.~(\ref{eqdf}), which makes the separation approximately
proportional to $A$ for a fixed energy. For
example, for He nuclei with $E=1$~EeV most of the separations are
smaller than 50~km, and hence within the range of existing or planned
detectors. 

\FIGURE{\epsfig{file=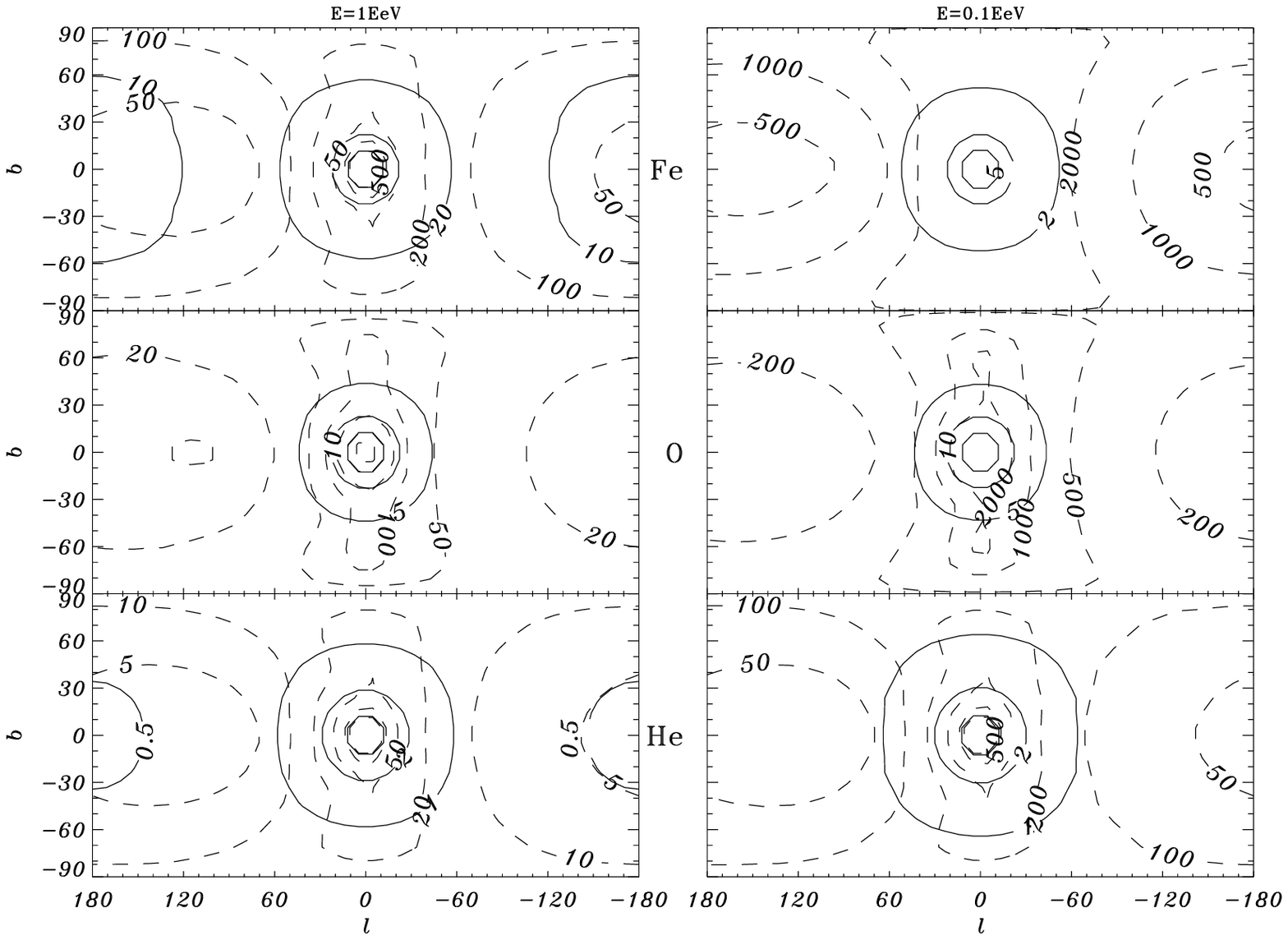,width=35em}\caption{Mean fragments 
separation (dashed curves) and
fragmentation
probability (solid lines) contours for total energy $E=0.1$~EeV and
$E=1$~EeV and for heavy (Fe), intermediate (O) and light (He) nuclei. The
separation contours are labelled by their values in km units  and the
probabilities in units of $10^{-6}$.\label{f2}}}

The solid contours in fig.~\ref{f2} correspond to the fragmentation
probabilities $\eta_{GZ}$ (and are labeled by $\eta_{GZ}\times
10^6$). Values of $\eta_{GZ}\sim 10^{-5}$--$10^{-4}$ are obtained 
 for Fe nuclei, while  $\eta_{GZ}\sim 10^{-6}$--$10^{-5}$
result for He and O.

In order to compute the rate of observation of these kind of events in
a detector covering a surface $S$, one must only consider showers
separated by distances smaller than $d_{max}\simeq \sqrt{S}$. The
rates will be a convolution of the CR flux, the fragmentation
probability and the fraction of GZ events with $d<d_{max}$,
$f_{d_{max}}(E)$. The areas (and thresholds) of the large extended air
shower detectors we will consider are 100~km$^2$ for AGASA~\cite{agasa}
($E_{th}\simeq 1$~EeV),  3000~km$^2$ for Auger~\cite{auger}
($E_{th}\simeq 1$~EeV), few$\times 10^4$~km$^2$ for HiRes~\cite{hires}
or the
proposed Japanese Telescope Array~\cite{ta} fluorescence detectors
($E_{th}\simeq 0.1$~EeV) and $\sim 10^6$~km$^2$ for the proposed 
OWL~\cite{owl} type satellite fluorescence detectors
($E_{th}\simeq 10$~EeV). 

Regarding the CR flux, above the knee in the
spectrum ($E_{CR}>3\times 10^{15}$~eV), the measured flux is
\beq
\Phi(E_{CR}>E)\simeq 47 \left({{\rm EeV}\over E}\right)^2[{\rm km^2\
yr\ sr}]^{-1}\,.
\eeq
Due to the steepness of the spectrum (d$\Phi/{\rm d}E\propto E^{-3}$),
we will just estimate the rates of events from CR with energy above a
given value $E$ as
\beq
Rate(E_{CR}>E)\simeq \Phi(E_{CR}>E)\eta_{GZ}(E)f_{d_{max}}(E)
S\epsilon_{dc}\Omega\,,
\eeq
where fluorescence detectors have typical duty cycles
$\epsilon_{dc}\simeq 0.1$ and $\Omega$ is the solid angle.

\FIGURE{\epsfig{file=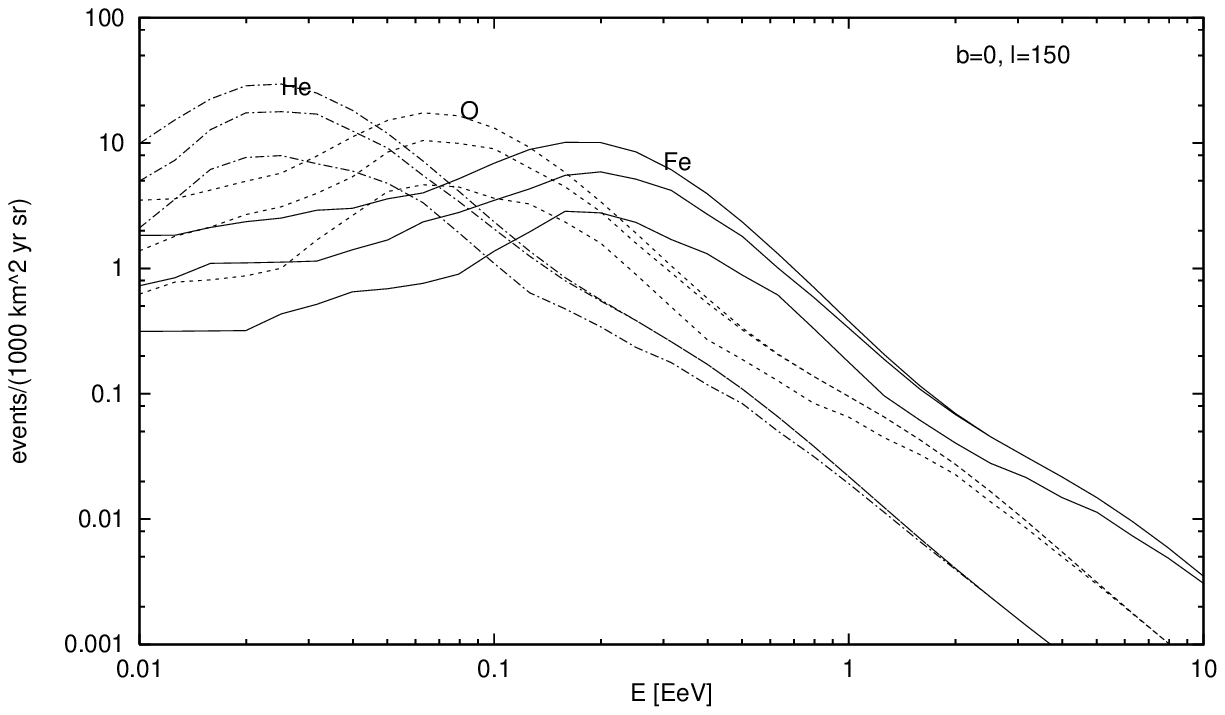}}
\EPSFIGURE{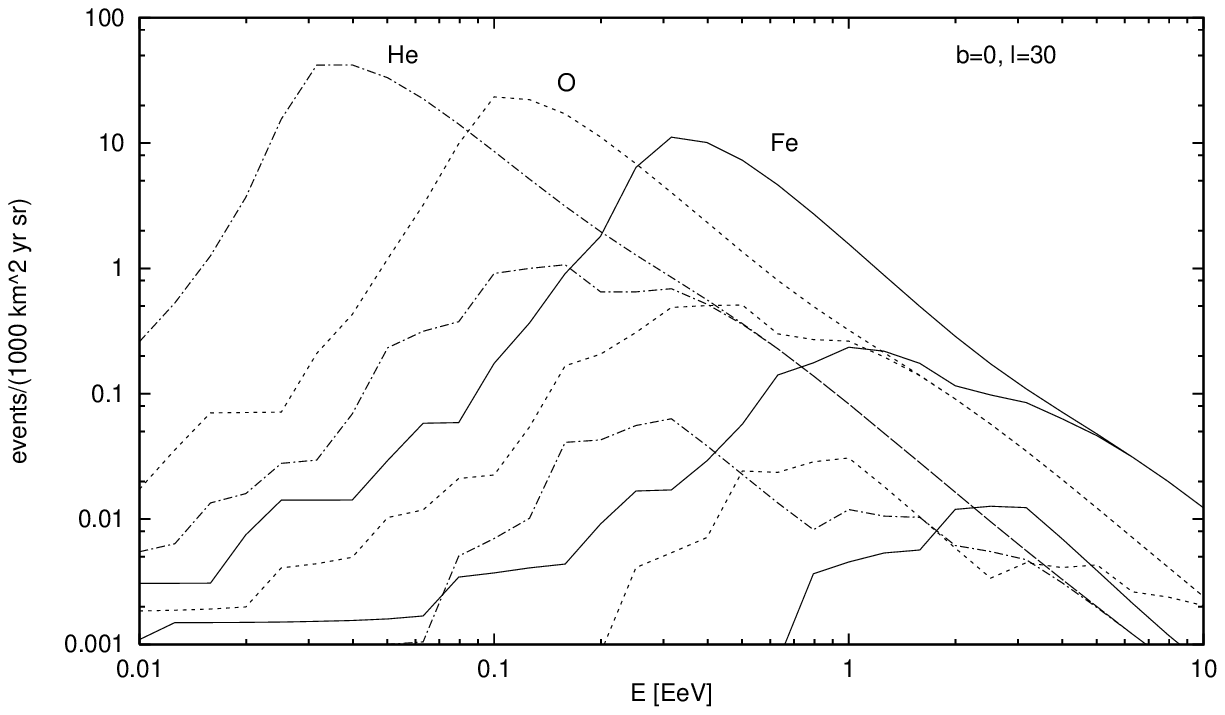}{Expected rates of events per year and sr for a 
detector area of
1000 km$^2$ and for separation between fragments smaller than 1000 km 
(upper curves), 100 km (middle curves) and 10 km (lower curves).
Solid lines correspond Fe, dashed lines to O and dot-dashed lines to
He. Fig.~3$a$ is for a night-time
direction ($b=0^\circ,\ \ell=150^\circ)$, while fig.~3\emph{b} for a direction
close to the sun ($b=0^\circ,\ \ell=30^\circ)$.\label{f3}}

In fig.~\ref{f3} we plot these rates as a function of the energy of the
parent CR, for the three nuclei and for maximum distances of 1000, 100
and 10~km. Fig.~\ref{f3}$a$ is for a night-time direction ($b=0^\circ,\ 
\ell=150^\circ)$
while fig.~\ref{f3}$b$ is for a direction close to the sun ($b=0^\circ,\ 
\ell=30^\circ)$.\footnote{Note that in the day-side, $|\ell |<90^\circ$,
for increasing $|b|$ the average deflections decrease,
enhancing the rates with respect to the $b=0$ results
of fig.~\ref{f3}. For the night-side the opposite will happen.}
It is clear that, if $d_{max} \lsim 100$~km, the maximum rates per 
unit coverage are attained
in the night side for energies $E \sim 10^{-2}$--1~EeV
(depending on the nucleus considered) and exceed one event per year
for coverages $S~\Omega~\epsilon_{dc}\gsim 10^3$~km$^2$~sr. Note that
the AGASA array has an order of magnitude smaller coverage, while 
the Auger array, with $S\simeq 3000~$km$^2$ looks in this sense 
more promising. However, to
be able to observe these events, low thresholds are required, and the
emitted nucleon will not fire a tank in Auger if its energy is below 
0.1~EeV. This means that to see both fragments, the parent CR should
have an energy larger than $0.1A~$EeV, which for incident Fe nuclei
is much larger than the typical threshold of the detector
($E_{th}\simeq 1$~EeV, corresponding to triggering with $\sim$~5
tanks). Hence, the rates for Auger are also quite small
($\lsim 10^{-1}$events/yr). The most promising experiments for
detecting these kind of events seem to be the large fluorescence
detectors in construction (HiRes) or planned (Telescope Array), 
since besides
having large areas ($\gsim~10^4$~km$^2$) they will have low thresholds
(0.1~EeV). Furthermore, it is even possible that if the heavier
fragment is above the threshold and triggers the detector, the lighter
fragment is seen even if it has an energy somewhat below 0.1~EeV.
With these arrays one will then expect few events per year for
energies below 1~EeV if the CR are indeed not mainly protons. Larger
detectors, as the proposed OWL and AIRWATCH satellites, have much
higher thresholds, $E_{th}~\simeq$~10EeV, and hence will have 
small rates ($\lsim~0.3$~events/yr).

Considering the day side ($|\ell|<90^\circ$), only ground arrays can
be employed, and looking at fig.~\ref{f3}$b$ we see that the rates are very
small for $d_{max}<100$~km. This is due to the fact that most fragment separations are
larger than the typical size of the detectors and become hence
unobservable.

\EPSFIGURE{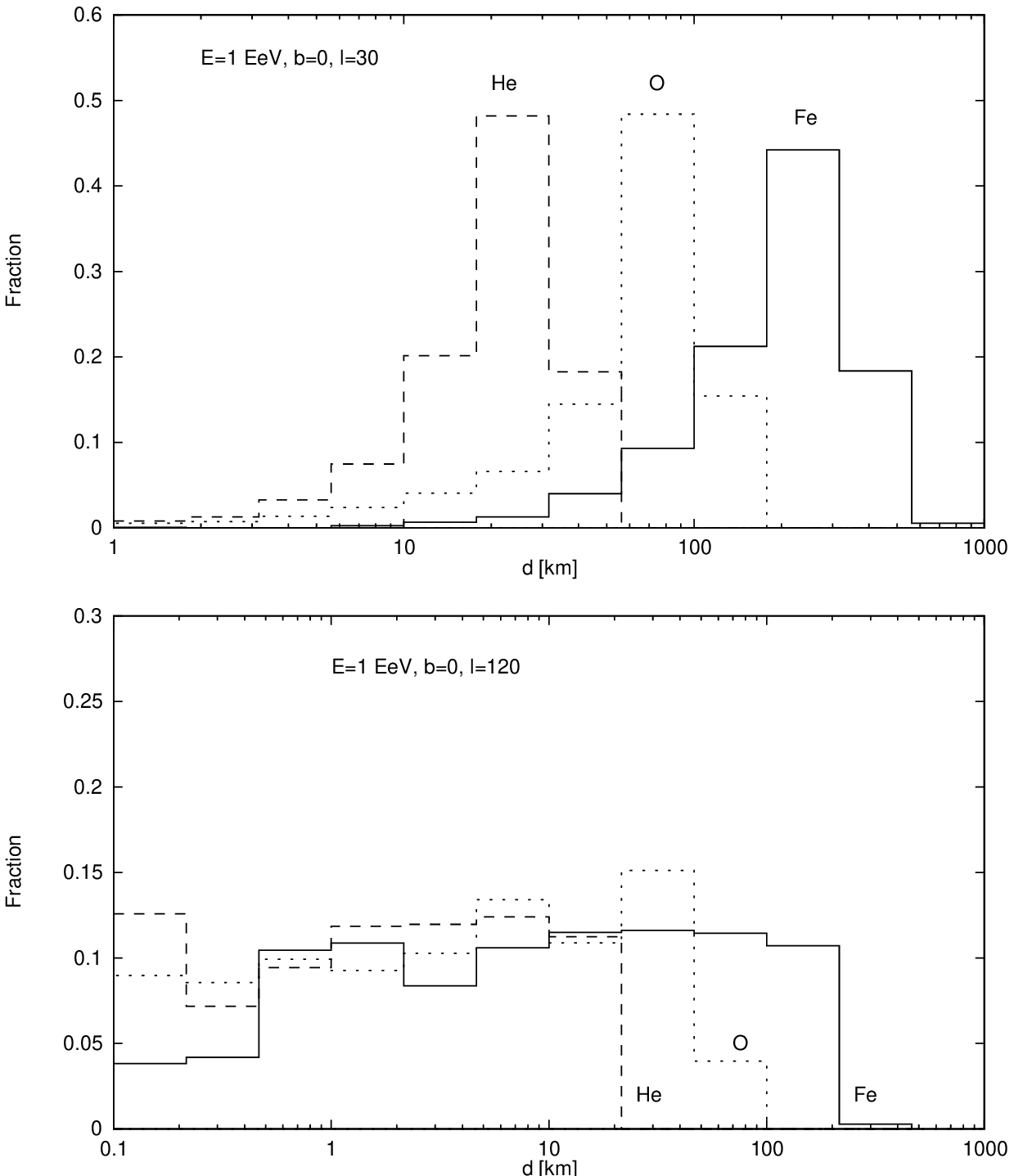}{Distribution of the separation 
between fragments for a total
energy of $E=1$~EeV and for the three nuclei  considered, 
Fe (solid line), O (dashed line) and He (dot-dashed line).
Fig.~4$a$ is for ($b=0^\circ,\ \ell=30^\circ)$, while fig.~4$b$ for 
 ($b=0^\circ,\ \ell=120^\circ)$. The histograms give the probability
that the separation falls in the range of distances corresponding to
each bin.\label{f4}}

In the case of a positive detection of two simultaneous showers, if a
single nucleon was emitted, the initial CR composition could be
determined as $A\simeq~(E_1+E_2)/E_1$, with $E_1$ the energy of the
less energetic shower and $E_2$ that of the most energetic one.
This procedure will not lead to a precise determination of $A$ if the
nucleon shower energy is close to the threshold, and hence poorly
determined, or if there was multiple nucleon emission, so that the
smaller shower is actually produced by an unknown number of nucleons.
There is  however additional information about the CR composition, encoded
in the separation between the detected fragments, which can be used 
to obtain a good determination of $A$ also in these cases. 
From fig.~\ref{f2} it is clear that, for a given initial energy and arrival
direction, the average separations  are quite sensitive to the CR
composition. However, the distributions of separations are quite flat 
(see fig.~\ref{f4}) due to the wide range of distances from the Earth at which
the photo-disintegration can occur (with the exception of the directions
close to the sun, e.g.\ fig.~\ref{f4}$a$, for which the most likely location for the
photo-disintegration interactions are for $\xi \simeq 1$~AU).
This can result  in different
nuclei producing similar separations if they disintegrate at different
distances.

We have found that this degeneracy can be lifted in many
cases by looking at the inclination of the showers (i.e.\ at the angle 
atan($d_b/d_\ell$)). In fig.~\ref{f5} we show the relations obtained between
the separation $d$ and the inclination angle, for $b=10^\circ$
(fig.~\ref{f5}$a$) and for $b=45^\circ$ (fig.~\ref{f5}$b$) and for different values of 
$\ell=30^\circ$, $60^\circ$, $90^\circ$, $120^\circ$, $150^\circ$, $180^\circ$, 
$240^\circ$ and $330^\circ$. One moves along the curves as the
photo-disintegration distance $\xi$ is varied. For the southern
hemisphere ($b<0^\circ$), the inclination angle changes sign. Since the
separations scale as $\gamma^{-1} \propto A/E$, what is plotted is 
$d\times(A/56)\times($EeV$/E$). 

Hence, knowing the total energy $E$,
one may infer $A$ by using plots like those in fig.~\ref{f5}. This is just done
by plotting the $d$ vs. angle curve for the arrival  direction 
($b,\ \ell$) of the
observed CR. Confronting then this curve with the measured separation
and inclination of the showers, and knowing the total energy $E$,
$A$ can be determined. One can see that for many directions
(e.g.\ $0^\circ<\ell<180^\circ$ in fig.~\ref{f5}$a$) the inclination of the
showers varies considerably as $\xi$ is varied, and hence the
inclination  contains the information on the distance to the
photo-disintegration point which allows the determination of  $A$. In
other directions (i.e.\ those leading to almost vertical curves) the 
resolving power of this method is instead  not good.
\EPSFIGURE[t]{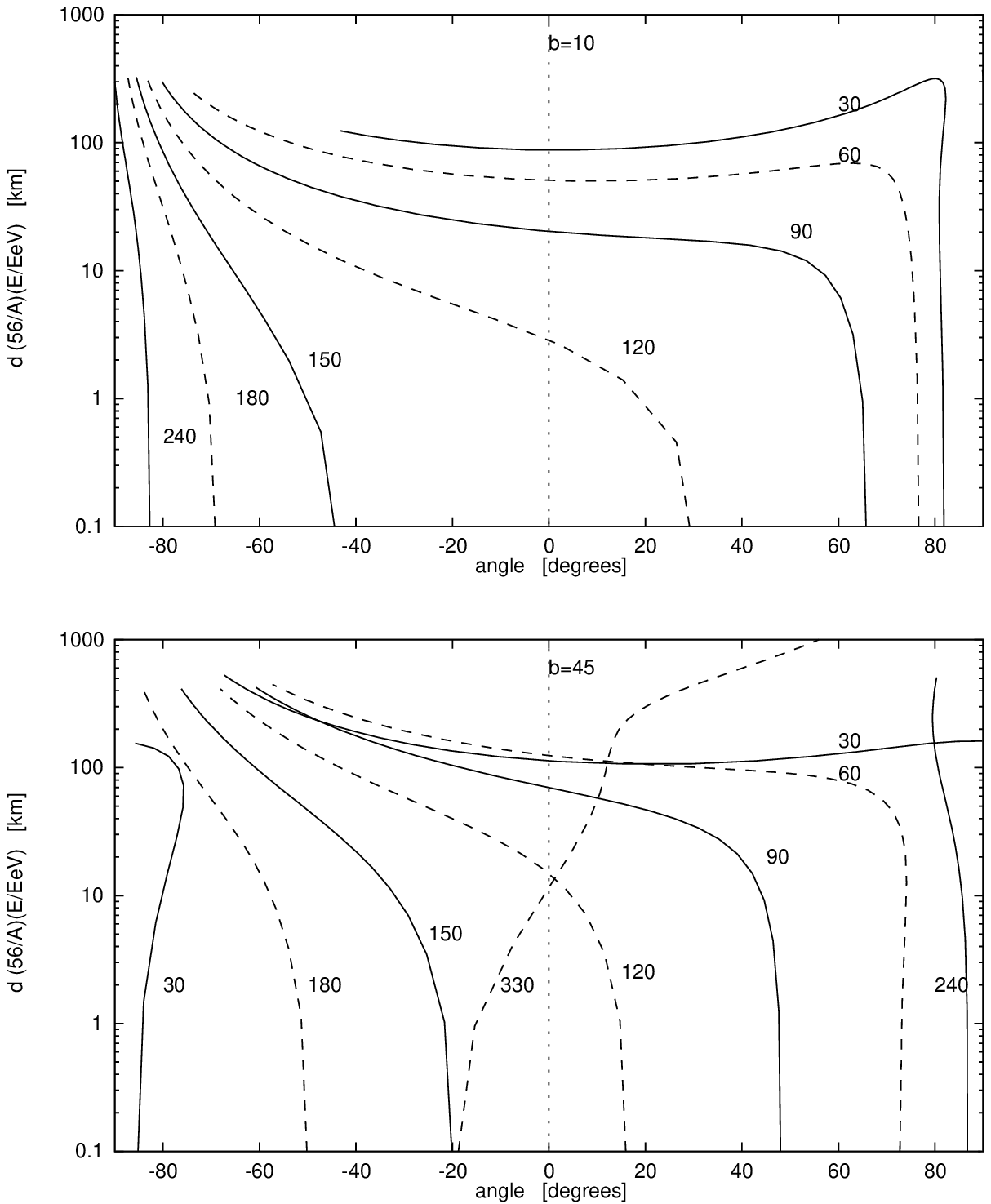}{Separation vs. inclination angle of the 
fragments for arrival
directions with $b= 10^\circ$ (fig.~5$a$) and $b= 45^\circ$  (fig.~5$b$),
and for the $\ell$ values indicated in each line.\label{f5}}

In summary, we have explored in detail the possibility of detecting double
shower events originating from the fragments produced in  the 
photo-disintegration of
a CR in a collision with a solar photon, for different compositions of the
primary CR. The best possibilities are for the future large area
fluorescence detectors, like HiRes and the Telescope Array, for which few
events per year are expected if the CR of energies larger than 0.1~EeV are
indeed nuclei. These will be rare, peculiar events, but it would be
interesting to  detect
them as they can give important information about the CR
composition at ultra-high energies. 
The proposed method of using the separation and inclination
of the showers to complement the analysis with the ratio of energies 
will also help to get a
more precise determination of the composition of the parent CR.

\acknowledgments

Work partially supported by CONICET, Argentina.

\providecommand\url[1]{\href{#1}{\tt #1}}

\end{document}